\documentclass[preprint,showpacs,preprintnumbers,amsmath,amssymb]{revtex4}

\usepackage{graphicx}
\usepackage{dcolumn}
\usepackage{bm}

\begin{document}

\preprint{Submission to Phys. Rev. B}

\title{
Magnetic excitations in the spin-5/2 antiferromagnetic trimer substance
SrMn$_3$P$_4$O$_{14}$
}

\author{Masashi Hase$^1$}
 \email{HASE.Masashi@nims.go.jp}
\author{Masaaki Matsuda$^{2,3}$}
\author{Koji Kaneko$^2$}
\author{Naoto Metoki$^2$}
\author{Kazuhisa Kakurai$^2$}
\author{Tao Yang$^4$}
\author{Rihong Cong$^4$}
\author{Jianhua Lin$^4$}
\author{Kiyoshi Ozawa$^1$}
\author{Hideaki Kitazawa$^1$}

\affiliation{%
${}^{1}$National Institute for Materials Science (NIMS), 
1-2-1 Sengen, Tsukuba, Ibaraki 305-0047, Japan \\
${}^{2}$Japan Atomic Energy Agency (JAEA), 
2-4 Shirakata Shirane, Tokai, Naka, Ibaraki 319-1195, Japan \\
${}^{3}$Oak Ridge National Laboratory (ORNL), 
Oak Ridge, TN 37831, USA\\
${}^{4}$College of Chemistry and Molecular Engineering, 
Peking University, Beijing 100871, People's Republic of China
}%

\date{\today}

\begin{abstract}

A quantum-mechanical 1/3 magnetization plateau and magnetic long-range order 
appear in the large-spin (5/2) substance SrMn$_3$P$_4$O$_{14}$. 
Magnetization results of SrMn$_3$P$_4$O$_{14}$ can be explained by 
the spin-5/2 isolated antiferromagnetic linear trimer with 
the intra-trimer interaction ($J_1$) value of 4.0 K. 
In the present study, to confirm the spin system, 
we performed inelastic neutron scattering (INS) experiments of SrMn$_3$P$_4$O$_{14}$ powders. 
We observed plural magnetic excitations. 
The peak positions are 0.46, 0.68, and 1.02 meV. 
Constant-$Q$-scan spectra at several $Q$ values (magnitude of the scattering vector) 
indicate that the dispersion is weak. 
The weak dispersion indicates that 
the excitations are transitions between discrete energy levels.  
Our INS results are consistent with results expected in the trimer model. 
We evaluated the $J_1$ value as 0.29 meV (3.4 K) without considering the other interactions. 

\end{abstract}

\pacs{75.40.Gb, 75.10.Jm, 75.47.Lx, 75.50.Ee}

\keywords{antiferromagnetic trimer, inelastic neutron scattering, magnetic excitation}
\maketitle

\section{Introduction}

Quantum-mechanical nature is sometimes apparent 
even in an ordered state of several low-dimensional spin systems formed by small spins. 
In the triangular antiferromagnet CsCuCl$_3$ with spin-1/2, 
a small jump was observed in the magnetization curve in the magnetic field parallel to the $c$ axis.\cite{Nojiri88} 
This jump was successfully explained as a spin flop process caused by quantum-mechanical effects.\cite{Nikuni93} 
In spin-gap systems with spin-1/2 such as 
the spin-Peierls system in CuGeO$_3$ \cite{Hase93a,Hase93c} 
and the two-leg ladder system in SrCu$_2$O$_3$,\cite{Dagotto92,Azuma94} 
antiferromagnetic long-range order (AF-LRO) appears when small amounts of impurities 
were doped.\cite{Hase93b,Hase95,Hase96a,Azuma97,Azuma98}, 
Nonetheless 
excitations originating in the singlet-triplet gap in the pure system 
were observed.\cite{Martin97}

It has not been studied adequately whether 
the quantum-mechanical nature can remain 
in an ordered state of spin systems formed by large spins.  
Quantum-mechanical nature is apparent in the ordered state
of the spin-5/2 substance SrMn$_3$P$_4$O$_{14}$.\cite{Yang08} 
The temperature $T$ dependence of the magnetic susceptibility 
indicates occurrence of a magnetic LRO below about 2.6 K. 
A 1/3 magnetization plateau in magnetization curves was observed at both 1.3 K and 4.2 K 
(below and above the transition temperature).\cite{Yang08,Hase09}
The plateau was observed in a powder sample. 
Therefore, the plateau can appear irrespective of the applied magnetic field direction. 
As a result, 
the plateau is a quantum-mechanical magnetization plateau 
generated by an energy gap in the magnetic excitation spectrum. 

It is important to determine the origins of 
the quantum-mechanical nature (magnetization plateau) in SrMn$_3$P$_4$O$_{14}$. 
We could explain well the magnetic-field $H$ and $T$ dependences of the magnetization 
using the spin-5/2 isolated linear trimer formed by the AF $J_1$ interaction ($J_1 = 4.0$ K) 
depicted as ellipses in Fig.1.\cite{Hase09} 
Therefore, we consider that 
the magnetization plateau originates in discrete energy levels of the AF trimer. 
If the AF trimer model is valid, 
we can observe magnetic excitations indicating the discrete energy levels 
even in powder samples. 
Consequently, we performed inelastic neutron scattering experiments of SrMn$_3$P$_4$O$_{14}$ powders. 

\begin{figure}
\begin{center}
\includegraphics[width=8cm]{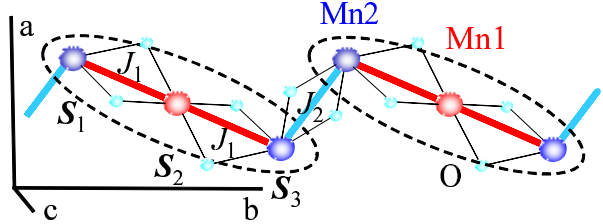}
\caption{
(Color online)
A schematic drawing of positions of Mn$^{2+}$ ions ($3d^5$) having localized spin 5/2 
in SrMn$_3$P$_4$O$_{14}$.\cite{Yang08} 
Two crystallographically independent Mn$^{2+}$ sites (Mn1 and Mn2) exist. 
Two kinds of short Mn-Mn bonds exist and have Mn-O-Mn paths. 
The Mn-Mn distances are 3.27 and 3.34 \AA \ at room temperature.
The exchange interaction parameters are respectively defined as $J_1$ and $J_2$. 
The dominant AF $J_1$ interactions form the spin trimers indicated by the ellipses. 
The Hamiltonian is expressed as ${\cal H} = J_1 (S_1 S_2 + S_2 S_3).$
The spin trimer can account for the magnetic-field and temperature dependences of the magnetization 
when $J_1 = 4.0$ K.  
Mn-Mn distances in the other bonds are more than 4.89 \AA. 
These bonds have no Mn-O-Mn paths. 
}
\end{center}
\end{figure}

\section{Methods of Experiments}

We synthesized single crystals of SrMn$_3$P$_4$O$_{14}$ 
under hydrothermal conditions at 473 K.\cite{Yang08}
Each crystal was small. 
We used pulverized crystals for inelastic neutron scattering (INS) measurements.  

We carried out INS measurements on 
the cold neutron triple-axis spectrometer LTAS installed at JRR-3M in JAEA. 
The final neutron energy was fixed at 2.6 meV. 
Higher-order beam contamination was effectively eliminated using 
a cooled Be filter before the sample.
The horizontal collimator sequence was guide-80'-Be-sample-120'-open. 
This setup yields an energy resolution of 0.1 meV 
(full width at half maximum, FWHM) at an energy transfer $\omega = 0$ meV. 
The resolution was determined from incoherent scattering of the sample. 
The powder sample of about 9 g was mounted in a ${}^4$He closed cycle refrigerator. 

\section{Results and discussion}

We performed all the INS measurements above the transition temperature. 
Circles in Fig. 2 show the $\omega$ dependence of the INS intensity
(constant-$Q$ scan spectra) around 5 K. 
The value of $Q$ is the magnitude of the scattering vector. 
Excitations are apparent between 0.5 and 1.5 meV. 
Two kinds of excitations with different peak positions seem to overlap each other. 
The spectra are almost independent of $Q$ except for difference in intensities. 
The weak $Q$ dependence indicates that the excitations are transitions between discrete energy levels. 
We consider that 
the intensities in the vicinity of 0 meV cannot be explained only by incoherent scattering 
because of $T$ dependence of spectra as shown later. 
Low-energy excitations exist. 

\begin{figure}
\begin{center}
\includegraphics[width=8cm]{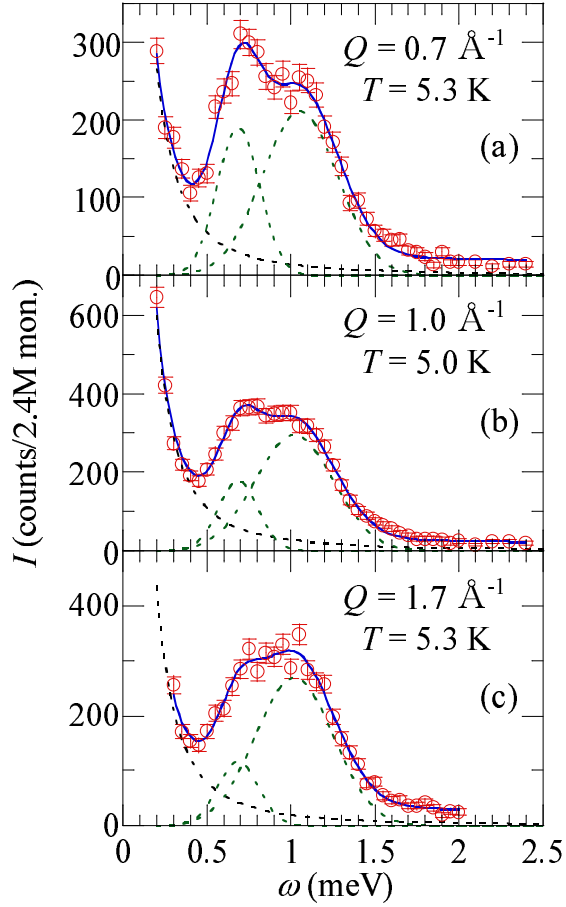}
\caption{
(Color online)
The inelastic neutron scattering intensity {\it vs.} energy transfer ($\omega$) 
[constant-$Q$ (magnitude of the scattering vector) scan spectra]
of SrMn$_3$P$_4$O$_{14}$ around 5 K (circles). 
The solid line represents the sum of 
two Gaussians and one Lorentzian (plus constant backgrounds). 
The dashed line indicates each Gaussian or Lorentzian. 
The values of the parameters are given in Table I. 
}
\end{center}
\end{figure}

We compared each spectrum above 0.2 meV in Fig. 2 with a sum of 
two Gaussians and one Lorentzian (plus constant backgrounds). 
\begin{equation}
I(\omega) = \frac{I_0 a_0}{\pi} \frac{1}{\omega^2 + a_0^2} +
\Sigma_{i} \frac{I_i}{\sqrt{\pi} a_i} \exp [- \frac{(\omega - \omega_i)^2}{a_i^2}] + I_{\rm BG}.
\end{equation}
Here the sum is from $i=1$ to 2.
The two Gaussians correspond to the excitations between 0.5 and 1.5 meV. 
The Lorentzian corresponds to the excitations in the vicinity of 0 meV. 
Each sum of two Gaussians and one Lorentzian (solid line) reproduces well the corresponding spectrum in Fig. 2. 
Obtained values of the fitting parameters are shown in Table I. 
The peak position in the spectrum at $Q=1.0$ \AA$^{-1}$ is 0.68 meV or 1.02 meV. 
The peak width (FWHM) is 0.28 meV and 0.55 meV for the 0.68 meV and 1.02 meV excitation, respectively. 
These widths are larger than the energy resolution of 0.1 meV at $\omega = 0$ meV, 
indicating existence of weak dispersion caused by inter-trimer interactions. 

\begin{table*}
\caption{\label{table1}
Values of the integrated intensity ($I_i$) and full width at half maximum (FWHM) of 
Lorentzian or Gaussian 
obtained from the fitting of Eq. (1) to experimental constant-$Q$ spectra of SrMn$_3$P$_4$O$_{14}$. 
FWHM is given as $2 a_0$ for Lorentzian or 
$2 \sqrt{\ln 2} a_i$ for Gaussian. 
In the fitting to the spectra at $Q =1.0$ \AA$^{-1}$, the value of FWHM of each Gaussian 
was obtained at 5.0 or 11.2 K and was fixed in the fitting at higher $T$. 
The values in parentheses indicate errors. 
}
\begin{ruledtabular}
\begin{tabular}{cccccccccc}
&& \multicolumn{2}{c}{0 meV} & \multicolumn{2}{c}{0.68(1) meV} & \multicolumn{2}{c}{1.02(2) meV} &\multicolumn{2}{c}{0.46(5) meV} \\
$Q$ & $T$ & $I_0$ & FWHM & $I_1$ & FWHM & $I_2$ & FWHM & $I_3$ & FWHM \\
\AA$^{-1}$ & K && meV && meV  && meV && meV \\
\hline
0.7 & 5.3 &   379(58) & 0.24(8) & 60(9) & 0.30(3) & 119(11) & 0.53(5) && \\
0.8 & 5.7 &   475(136) & 0.31(10) & 58(8) & 0.34(4) & 159(21) & 0.62(10) && \\
1.0 & 5.0 & 1091(33) & 0.16(4) & 52(9) & 0.28(3) & 172(13) & 0.55(4) && \\
1.2 & 5.4 & 1123(34) & 0.16(4) & 32(9) & 0.25(5) & 190(14) & 0.53(4) && \\
1.5 & 5.1 &   968(54) & 0.18(3) & 31(6) & 0.23(3) & 177(13) & 0.54(3) && \\
1.7 & 5.3 &   795(51) & 0.16(4) & 36(13) & 0.29(6) & 157(13) & 0.55(6) && \\
\hline
1.0 &   5.0 & 1091(33) & 0.16(4) & 52(9) & 0.28(3) & 172(13) & 0.55(4) && \\
      & 11.2 &   562(296) & 0.43(20) & 36(6) & 0.28 & 156(12) & 0.55 & 17(12) & 0.25(0.14) \\
      & 15.6 &   515(78) & 0.53(7) & 31(6) & 0.28 & 141(12) & 0.55 & 17(11) & 0.25 \\
      & 20.3 &   518(179) & 0.56(16) & 29(6) & 0.28 & 117(11) & 0.55 & 17(10) & 0.25 \\
\end{tabular}
\end{ruledtabular}
\end{table*}

Circles in Fig. 3 show constant-$Q$ scan spectra at $Q=1.0$ \AA$^{-1}$.  
The  0.68 meV and 1.02 meV excitations are also seen at 11.2 K. 
Intensities around 0.5 meV are larger at 11.2 K than at 5.0 K, 
suggesting appearance of another transition. 
Therefore, we compared the spectrum above 0.2 meV at 11.2 K with a sum of 
three Gaussians and one Lorentzian (plus constant backgrounds) 
given in Eq. (1) with $i = 1$ to 3. 
To reduce variable parameters, we assumed that 
the peak position (0 meV) of the Lorentzian, and  
the peak positions (0.68 and 1.02 meV) and widths of the two Gaussians were constant.
We used the values obtained at 5.0 K. 
This assumption is reasonable for transitions between discrete energy levels. 
The sum of the three Gaussians and one Lorentzian (solid line) reproduces well the spectrum at 11.2 K. 
The peak position of the third Gaussian is 0.46 meV. 
The peak width (FWHM) is 0.25 meV and 
is larger than the energy resolution of 0.1 meV at $\omega = 0$ meV.

\begin{figure}
\begin{center}
\includegraphics[width=8cm]{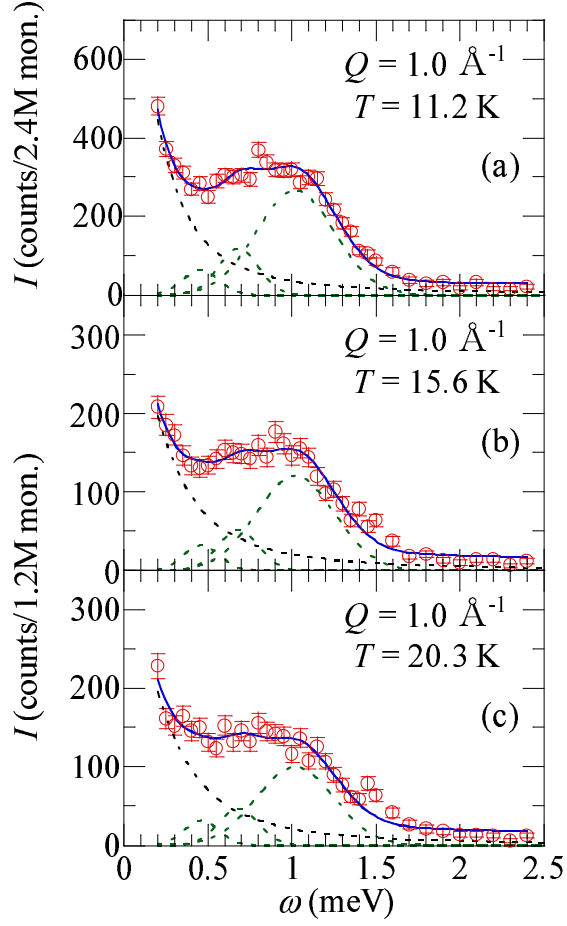}
\caption{
(Color online)
Constant-$Q$ scan spectra of 
SrMn$_3$P$_4$O$_{14}$ at $Q=1.0$ \AA$^{-1}$ (circles). 
The solid line represents the sum of 
three Gaussians and one Lorentzian (plus constant backgrounds). 
The dashed line indicates each Gaussian or Lorentzian. 
The values of the parameters are given in Table I. 
}
\end{center}
\end{figure}

The spectra at 15.6 K and 20.3 K shown respectively in Figs. 3(b) and (c) 
resemble the spectrum at 11.2 K. 
Therefore, we compared the spectrum above 0.2 meV at 15.6 K or 20.3 K 
with a sum of three Gaussians and one Lorentzian (plus constant backgrounds) 
given in Eq. (1) with $i = 1$ to 3.  
In the fitting, we assumed that 
the peak position of the Lorentzian, and  
the peak positions and widths of the three Gaussians were constant.  
The sum (solid line) reproduces well each experimental spectrum. 
Spectra are featureless above 30 K. 
We did not compare the spectra with calculated curves. 
The integrated intensity between 0.5 and 1.5 meV decreases slightly on heating. 
The Bose factor proportional to phonon intensity at 0.68 meV and 20.3 K, on the other hand, is about 9 times as large as 
that at 0.68 meV and 5.0 K. 
Therefore, contribution of phonon is small enough and 
magnetic excitations are dominant between 0.5 and 1.5 meV. 

We examined whether the spin-5/2 AF trimer model with $J_1 = 4.0$ K 
can account for the observed excitations. 
Figures 4(a) depicts 
a schematic drawing of low-lying energy levels.\cite{Hase09} 
The following selection rules of transitions are derived theoretically.\cite{Furrer79}
\begin{equation}
\Delta S = 0, \pm 1, \ \  \Delta M = 0, \pm 1, \ \ {\rm and} \ \ \Delta S_{13} = 0, \pm 1.
\end{equation}
$S_i (i=1, 2, 3)$ is the spin operator in the trimer. 
$S$ and $S_{13}$ is defined as $S_1 + S_2 + S_3$ and $S_1 + S_3$, respectively. 
$M$ is the $z$ component of $S$. 
Arrows in Fig. 4(a) indicate allowed transitions from 
the ground state (GS), first-excited state (1ES) or second-excited states (2ES).  
In our experimental setup, 
we can observe transitions with an energy difference $\Delta \epsilon$ up to 8
when $J_1 = 4.0$ K. 
Here, $\epsilon$ is defined as $E/J_1$ ($E$: eigen energy).

\begin{figure}
\begin{center}
\includegraphics[width=8cm]{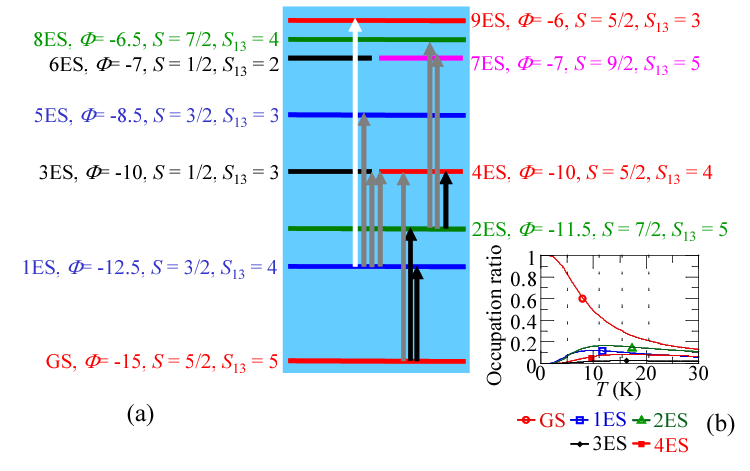}
\caption{
(Color online)
(a) 
A schematic drawing of low-lying energy levels in 
the spin-5/2 AF linear trimer 
[ground state (GS) and excited states (ESs)].\cite{Hase09} 
The parameters $\epsilon \equiv E/J_1$ and $S$ indicate 
the eigen energy and the total spin, respectively. 
$S_{13}$ is defined in the text. 
To distinguish two degenerate eigen states with $\epsilon = -10$, 
we name the two states 3ES and 4ES. 
To distinguish two degenerate eigen states with $\epsilon = -7$, 
we name the two states 6ES and 7ES. 
The arrows indicate allowed transitions from GS, 1ES or 2ES  
with an energy difference $\Delta \epsilon$ up to 8.  
We observed the three black-arrow transitions. 
The gray-arrow transitions may exist. 
We could not detect the white-arrow transition. 
(b) 
The temperature dependence of calculated occupation ratio 
of the five low-lying energy levels when $J_1 = 4.0$ K. 
}
\end{center}
\end{figure}

Figure 4(b) depicts 
$T$ dependence of calculated occupation ratio of the five low-lying energy levels. 
An inelastic neutron scattering intensity strongly depends on the occupation ratio. 
From Fig. 4(b), we know that excitations from GS are dominant around 5 K. 
We considered that 
the 0.68 and 1.02 meV excitations correspond to transitions 
from GS to 1ES and 2ES, respectively, indicated by black arrows.  
The respective energy differences are $2.5 J_1$ and $3.5 J_1$. 
The value of $J_1$ is evaluated as 0.27 meV (3.2 K) or 0.29 meV (3.4 K). 
These values are slightly smaller than the value determined in 
the magnetization results ($J_1 = 4.0$ K). 
Excitations from 1ES or 2ES are also expected at 11.2 K. 
We considered that 
the 0.46 meV excitation corresponds to the transition 
from 2ES to 4ES indicated by a black arrow.  
The energy difference is $1.5 J_1$. 
The value of $J_1$ is evaluated as 0.30 meV (3.5 K) and 
is close to the values evaluated from the other two transitions. 

We examined whether we observed all the allowed transitions 
that are possible in our experimental setup. 
As was described, we observed the three black-arrow transitions. 
The gray-arrow transitions may exist. 
However, we could not prove the existence of the gray-arrow transitions. 
Energy differences of some gray-arrow transitions ($2.5 J_1$ and $4 J_1$)
are the same as or close to the energy differences of black-arrow transitions 
($2.5 J_1$ and $3.5 J_1$). 
Therefore, we could not extract contribution of the gray-arrow transitions 
from the experimental results. 
Energy differences of the other gray-arrow transitions are 
$4.5 J_1 = 1.3$ meV and $5 J_1 = 1.4$ meV when $J_1 = 0.29$ meV. 
Small INS intensities are seen around these energies in Figs. 2 and 3. 
In our analyses, the small intensities correspond to the tail of the 1.02 meV excitation. 
However, the peak width is larger in the 1.02 meV excitation (FWHM = 0.55 meV) 
than in the 0.46 meV excitation (FWHM = 0.25 meV) or the 0.68 meV excitation (FWHM = 0.28 meV).  
The gray-arrow transitions may exist in the tail. 
We could not detect the white-arrow transition. 
We do not have theoretical INS intensities. 
Therefore, we could not determine the reason why we could not detect the white-arrow transition. 
The INS intensity of the white-arrow transition may be very small. 
Transitions from 3ES or higher excited states must exist at 11.2 K and higher $T$. 
However, we could not prove existence of these transitions 
because of the same reason for the gray-arrow and white-arrow transitions in Fig. 4(a) 

Figure 5 shows the $T$ dependence of the integrated intensity 
of the 0.46, 0.68, and 1.02 meV excitations. 
The integrated intensity of the 0.46 meV excitation is nearly independent of $T$. 
The integrated intensity of the 0.68 or 1.02 meV excitation 
gradually decreases with increasing $T$. 
As is shown in Fig. 4(a), 
several transitions are expected to exist. 
Therefore, the $T$ dependence of the integrated intensity in Fig. 5 
cannot be compared directly with the occupations ratios in Fig. 4(b). 

\begin{figure}
\begin{center}
\includegraphics[width=8cm]{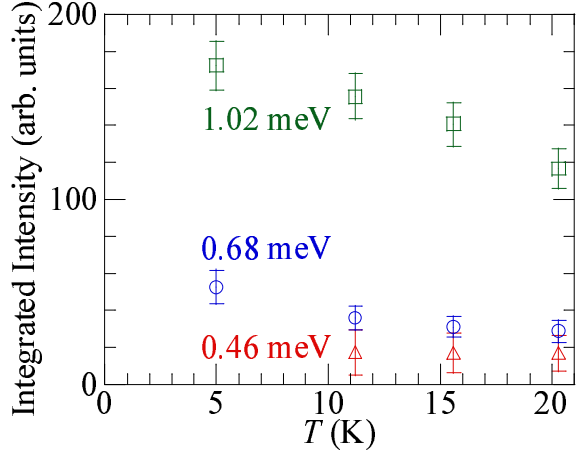}
\caption{
(Color online)
The temperature dependence of the integrated intensity 
of the 0.46, 0.68, and 1.02 meV excitations in SrMn$_3$P$_4$O$_{14}$ at $Q=1.0$ \AA$^{-1}$. 
}
\end{center}
\end{figure}

The $Q$ dependence of the INS intensity in the AF trimer is given 
in the following formula.\cite{Furrer79,Matsuda05,Podlesnyak07}
\begin{equation}
I(Q) = A_1 f(Q)^2 [1 - \sin(3.27Q)/(3.27Q)] + A_2 f(Q)^2 [1 - \sin(6.54Q)/(6.54Q)]. 
\end{equation}
The values 3.27 and 6.54 indicates the Mn1-Mn2 and Mn2-Mn2 length in the AF trimer, respectively. 
The function $f(Q)$ is the magnetic form factor of Mn$^{2+}$ ions.\cite{InternationalTable} 
The coefficients $A_1$ and $A_2$ depend on two eigen states between which the transition occurs. 
The coefficients are not derived theoretically. 
Figure 6(a) represents the two terms in Eq. (3). 
Circles in Figs. 6(b) - (d) show constant-$\omega$ scan spectra of SrMn$_3$P$_4$O$_{14}$. 
The lines indicate the first term in Eq. (3) plus constant backgrounds.
The INS intensity in the vicinity of $\omega = 2$ meV is small and 
almost independent of $\omega$, $Q$, and $T$.  
Therefore, we used the intensity at $\omega = 2$ meV for 
the value of constant backgrounds. 
The each line is consistent with the corresponding constant-$\omega$ scan spectrum. 
If $A_1$ is much larger than $A_2$ in the observed transitions, 
this consistency indicates that the AF trimer model can explain the experimental $I(Q)$. 

\begin{figure}
\begin{center}
\includegraphics[width=8cm]{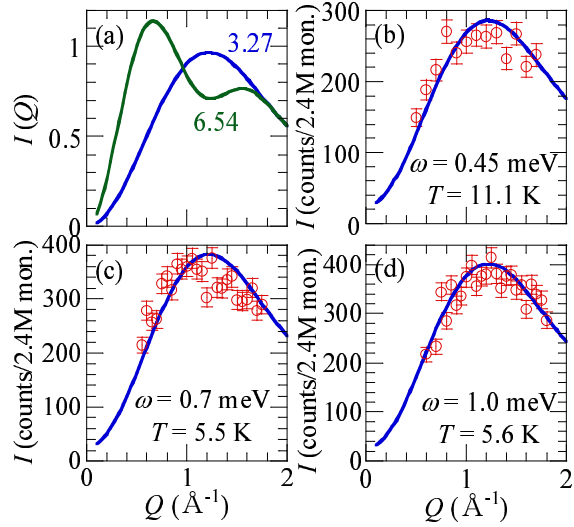}
\caption{
(Color online)
(a) 
The $Q$ dependence of the calculated INS intensity in 
the spin-5/2 AF linear trimer.  
(b)-(d) 
The inelastic neutron scattering intensity {\it vs.} $Q$
(constant--$\omega$ scan spectra) of 
SrMn$_3$P$_4$O$_{14}$ (circles). 
The line indicates the first term in Eq. (3). 
}
\end{center}
\end{figure}

We comment on inter-trimer interactions. 
The dispersion relation of magnetic excitations was calculated 
in spin dimers with weak inter-dimer interactions  
using random phase approximation.\cite{Leuenberger84}
A similar dispersion relation was inferred in interacting spin tetramers.\cite{Hafliger09}
According to the results, we speculate that 
the following  dispersion relation may be applicable to 
spin trimers with weak inter-trimer interactions. 
\begin{equation}
\omega_{{\bf q}=(h, k, l)} = \sqrt{ \Delta^2 + \alpha \Delta J({\bf q}) R(T)}.
\end{equation}
Here, 
$\Delta$ is an energy difference between ground and excited states. 
$\alpha$ is a coefficient derived from transition matrix elements. 
The value of $\alpha$ is 
2 for the spin-1/2 dimer \cite{Sasago97} or 
5 for the spin-3/2 dimer.\cite{Leuenberger84}
$J({\bf q})$ is a Fourier transform of inter-trimer interactions. 
$R(T)$ is a difference in thermal populations of ground and excited states. 
We consider that the dominant inter-trimer interaction is the $J_2$ interaction. 
$J({\bf q})$ is expressed approximately as $2 J_2 cos (2 \pi k)$.
We assume that excitation energies are the same at 0 and 5.0 K. 
The excitation energy at 0 K at the bottom of the dispersion $\omega_{\rm b}$, 
where the INS intensity is the strongest, 
is expressed as follows. 
\begin{equation}
\omega_{\rm b} = \sqrt{ \Delta^2 - 2 \alpha \Delta J_2}.
\end{equation} 
We assume that only the $J_2$ interaction is the origin of the difference between 
the expected excitation energy $2.5 J_1 = 10$ K and 
experimental excitation energy 0.68 meV = 7.9 K in the transition between GS and 1ES. 
Using $\Delta = 10$ K and $\omega_{\rm b} = 7.9$ K,
we obtained $\alpha J_2 = 1.9$ K. 
If $\alpha$ is large, a $J_2$ value can be small enough in comparison with the $J_1$ value. 

We observed magnetic excitations that are consistent with 
excitations expected in the spin-5/2 AF trimer. 
Therefore, the discrete energy levels of the AF trimer are the origins of 
the quantum-mechanical nature (magnetization plateau) in SrMn$_3$P$_4$O$_{14}$.  
In the strict sense, the energy difference between GS (S=5/2) and 2ES (S=7/2) 
generates the magnetization plateau. 
The magnetization plateau appears even in the ordered state.\cite{Yang08,Hase09} 
The property of the cluster (trimer in this case) can remain in the ordered state. 
The total spin of the ground state of the AF trimer is finite (5/2). 
Therefore, we consider that the magnetic LRO is stabilized by the $J_1$ and 
weak three-dimensional inter-trimer interactions. 
Several cluster substances can maintain their cluster properties in their ordered states. 
For example, 
the 1/2 quantum-mechanical magnetization plateau is generated by discrete energy levels of 
a spin-1/2 tetramer in Cu$_2$CdB$_2$O$_6$.\cite{Hase05,Hase09b}
The plateau remains in the ordered state. 
With the aid of other researchers, 
some of the present authors determined the magnetic structure 
below the transition temperature of $T_{\rm N} = 2.2(1)$ K 
using neutron powder diffraction data.\cite{Hase11} 
The magnetic structure has a long-range period. 
We are now considering the origin of the long-range period. 
We will report 
details of the magnetic structure in a subsequent paper. 

We comment on the INS intensity in the vicinity of 0 meV. 
Figure 7 shows constant-$Q$ scan spectra at $Q=1.0$ \AA$^{-1}$ below 0.5 meV. 
As was described, the increase of the intensity around 0.4 meV is caused by the 0.46 meV excitation. 
The intensity between 0.1 and 0.2 meV decreases with increasing $T$. 
This temperature dependence cannot be explained by incoherent scattering or phonon. 
In addition, we observed diffuse scattering between $2 \theta = 15$ and $40^{\circ}$ 
in neutron powder diffraction patterns (wavelength $\lambda=2.458$~\AA).\cite{Hase11} 
This $2 \theta$ range corresponds to $Q = 0.7$ to 1.7 \AA$^{-1}$ in the present INS experiments. 
The shape of the diffuse scattering resembles one-dimensional or two-dimensional Bragg scattering 
with a cutoff at low $Q$ and long tail at large $Q$.  
The integrated intensity of the diffuse scattering shows a maximum in the vicinity of $T_{\rm N}$. 
Several magnetic reflections appear below $T_{\rm N}$ between $2 \theta = 15$ and $40^{\circ}$. 
Therefore, the origin of the diffuse scattering is magnetic. 
Consequently, magnetic excitations exist in the vicinity of 0 meV. 
The magnetic excitations cannot be explained by transitions between energy levels in the trimer. 
Spin fluctuation in the ground state generates the magnetic excitations. 
Therefore, we used the Lorentzian with the 0 meV peak 
in the fitting of the constant-$Q$ scan spectra in Figs. 2 and 3. 
In future studies, we will perform INS measurements of SrMn$_3$P$_4$O$_{14}$ 
in the ordered state. 
Anisotropy of the Mn$^{2+}$ spins is small.\cite{Hase09} 
Therefore, a gap of spin-wave excitations is small.  
We may observe spin-wave excitations in the vicinity of 0 meV in addition to the trimer excitations.  

\begin{figure}
\begin{center}
\includegraphics[width=8cm]{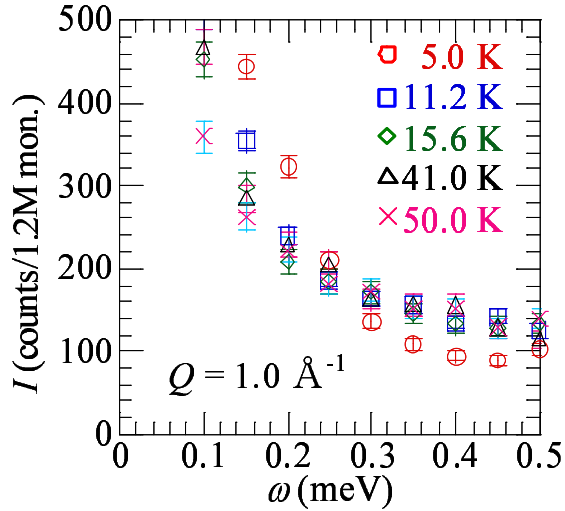}
\caption{
(Color online)
Constant-$Q$ scan spectra of 
SrMn$_3$P$_4$O$_{14}$ at $Q=1.0$ \AA$^{-1}$. 
}
\end{center}
\end{figure}

\section{Summary}

In order to confirm the spin system, 
we performed inelastic neutron scattering (INS) experiments of powders of 
the spin-5/2 antiferromagnetic trimer substance SrMn$_3$P$_4$O$_{14}$. 
We observed plural magnetic excitations. 
The peak positions are 0.46, 0.68, and 1.02 meV. 
The weak $Q$ dependence of constant-$Q$-scan spectra
indicates that the excitations are transitions between discrete energy levels. 
The experimental results are consistent with results  
expected in the trimer model with the intra-trimer interaction value of 0.29 meV (3.4 K) 
without considering the other interactions. 

\begin{acknowledgments}

We are grateful 
to T. Masuda for invaluable discussion. 
The neutron scattering experiments were carried out 
in the framework of JAEA Users' Program and 
within the NIMS-RIKEN-JAEA Cooperative Research Program on Quantum Beam Science and Technology. 
This work was partially supported 
by grants from NIMS.  

\end{acknowledgments}


\end{document}